\documentclass[12pt,letterpaper]{article}

\usepackage{cvpr}
\usepackage{times}
\usepackage{epsfig}
\usepackage{graphicx}
\usepackage{amsmath}
\usepackage{amssymb}
\usepackage{longtable}
\usepackage{paralist}
\usepackage{multirow,epstopdf,subfigure}
\usepackage{enumitem}
\usepackage{algorithm}
\usepackage{algorithmic}
\usepackage{bm}
\usepackage{color} 

\usepackage[breaklinks=true,bookmarks=false]{hyperref}

\cvprfinalcopy 

\def\cvprPaperID{****} 

\begin{document}

\title{Noise2Blur: Online Noise Extraction and Denoising}

\author{Huangxing Lin\textsuperscript{1}, \ Weihong Zeng\textsuperscript{1},\  Xinghao Ding\textsuperscript{1*},\ Xueyang Fu\textsuperscript{2}, \  Yue Huang\textsuperscript{1},\ John Paisley\textsuperscript{3}\\
\normalsize \textsuperscript{1}Xiamen University,
\normalsize \textsuperscript{2}University of Science and Technology of China, China\\
\normalsize \textsuperscript{3}Columbia University, New York, NY, USA\\
{\small \tt  $\{hxlin, zengwh \}$@stu.xmu.edu.cn, dxh@xmu.edu.cn,}\\
{\small \tt xyfu@ustc.edu.cn, huangyue05@gmail.com, jpaisley@columbia.edu}
}

\maketitle


\begin{abstract}
	We propose a new framework called Noise2Blur (N2B) for training robust image denoising models without pre-collected paired noisy/ clean images. The training of the model requires only some (or even one) noisy images, some random unpaired clean images, and noise-free but blurred labels obtained by predefined filtering of the noisy images. The N2B model consists of two parts: a denoising network and a noise extraction network. First, the noise extraction network learns to output a noise map using the noise information from the denoising network under the guidence of the blurred labels. Then, the noise map is added to a clean image to generate a new ``noisy/clean'' image pair. Using the new image pair, the denoising network learns to generate clean and high-quality images from noisy observations. These two networks are trained simultaneously and mutually aid each other to learn the mappings of noise to clean/blur. Experiments on several denoising tasks show that the denoising performance of N2B is close to that of other denoising CNNs trained with pre-collected paired data.
	
\end{abstract}

\section{Introduction}

Image denoising, which aims to restore a high-quality image from its degraded observation, is a fundamental problem in image processing. In many imaging systems \cite{eo2018kiki, lee1999polarimetric,yousuf2011new, jeyalakshmi2010modified}, image noise comes from multiple sources, such as capturing instruments, data transmission media or subsequent post-processing. Such complicated generation processes makes it difficult to estimate the latent noise model and recover the clean image from the noisy observation.

A large variety of denoising algorithms have been developed to deal with image noise. Traditional denoising methods (\emph{e.g.} BM3D \cite{dabov2007image}, WNNM \cite{gu2014weighted}) exploit a property of the noise or image structure to help denoising. These methods require accurate image model definitions, thus performance is limited in real-world cases. Modern denoising methods often employ convolutional neural networks (CNNs) to learn the mapping function from noise to clean on a large collection of noisy/clean image pairs. The performance of CNN denoisers are highly dependent on whether the distributions of training noise and test noise are well matched. Since pairs of real noisy/clean images are difficult to obtain, CNN denoisers are mostly trained on synthesized data. In addition, the real noise degradation process is usually complex or unknown, so that the synthesized noise distribution can deviate severely from the real noise distribution. As a result, CNN denoisers are easily over-fitted to the specific synthetic noise (\emph{e.g.} Gaussian noise, Poisson noise) and generalize poorly to the real-world noisy images.

In this paper, we propose Noise2Blur (N2B), a training scheme that overcomes the above problems. The training of N2B model does not need access to estimation of noise and pre-collected paired data. It only requires some unpaired noisy images and clean images, which is easy to implement in most practical applications. Although the images we have are unpaired, we can extract information from them to generate supervision for the denoising process. Our N2B model consists of two subnetworks, \emph{i.e.} denoising and noise extraction. The noisy inputs are first transformed into noise-free but blurred images by general filtering techniques (\emph{e.g.} Gaussian filter, median filter). We use the noisy/blurred image pair to guide the noise extraction network to roughly extract noise from its input (``noise-to-blur''). The denoising network is then trained using a new noisy/clean image pair obtained by adding the extracted noise to a random clean image (``noise-to-clean''). With a simple gradient interruption operation, the denoising network eventually converges to the ``noise-to-clean'' objective, while the noise extraction network learns to finely extract the noise. On the other hand, the training of N2B model has no requirement on the number of noisy images; even if only one noisy image of size $512\times512$ is available, a denosing network with strong generalization can be trained. Through experiments on several datasets, we demonstrate the effectiveness of N2B. Although we train with unpaired data, the denoising performance of the N2B model is close to that of other denoising CNNs trained with pre-collected paired data.

\begin{figure}[t]
	\centering
	\includegraphics[width=0.73in]{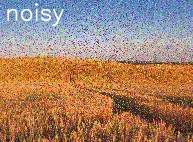}	
	\includegraphics[width=0.73in]{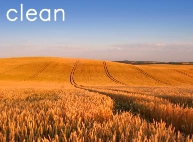}
	\includegraphics[width=0.73in]{1-noise1.jpg}	
	\includegraphics[width=0.73in]{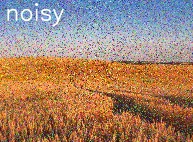}	
	\includegraphics[width=0.73in]{1-noise1.jpg}	
	\includegraphics[width=0.73in]{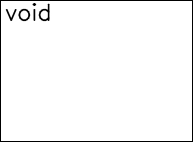}
	\includegraphics[width=0.73in]{1-noise1.jpg}
	\includegraphics[width=0.73in]{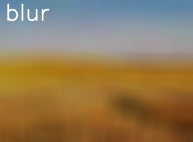}\\
	\vspace{-0.07in}
	\subfigure[\tiny{SSIM, PSNR $|$0.981, 39.34dB}]{\includegraphics[width=1.5in]{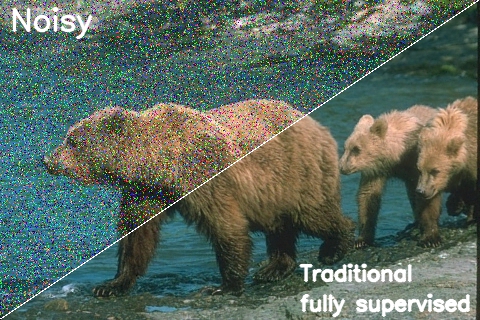}}
	\subfigure[\scriptsize{0.976, 38.25dB}]{\includegraphics[width=1.5in]{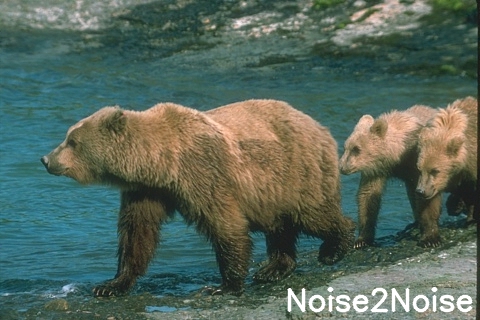}}	
	\subfigure[\scriptsize{0.828, 28.95dB}]{\includegraphics[width=1.5in]{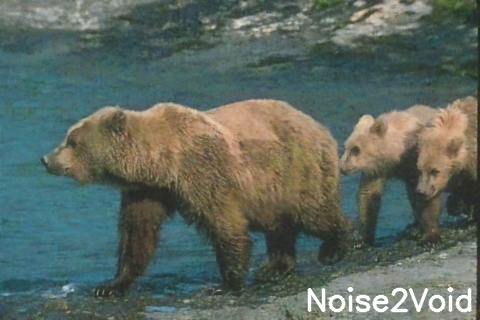}}
	\subfigure[\scriptsize{0.968, 37.26dB}]{\includegraphics[width=1.5in]{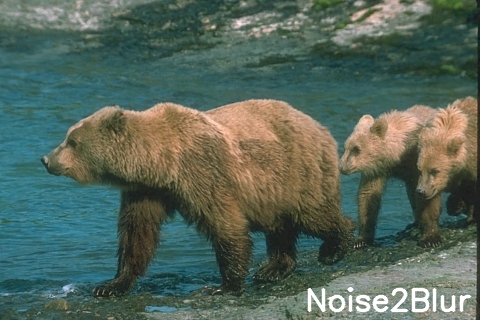}}\\

	\caption{Comparison of different training schemes. (a) Traditionally, the training of denoising network requires a large amount of paired noisy/clean images. (b) Noise2Noise \cite{lehtinen2018noise2noise} trains the network using pairs of independent noisy measurements of the same target. (c) Noise2Void \cite{krull2019noise2void} uses just individual noisy images as training data. (d) Our Noise2Blur needs some unpaired noise and clean images, as well as blurred counterparts of the noisy images to generate supervision.}
	\label{fig1}
\end{figure}

\section{Related work}

In general, image denoising methods can be categorized into model-based methods and learning-based methods.

Model-based methods, which contains the most traditional image denoising methods, exploit the image priors about the underlying signal, such as self-similarity \cite{dabov2007image}, smoothness \cite{buades2005review} or sparsity \cite{elad2006image, mairal2007sparse}. Along this line, anisotropic diffusion \cite{perona1990scale}, total variation denoising \cite{rudin1992nonlinear} and wavelet coring \cite{simoncelli1996noise} use the particular statistical regularities of images to recorve a clean image from noisy input. Later, the non-local self-similarity (NSS) prior \cite{dabov2007image, buades2005non, mairal2009non, lefkimmiatis2017non}, which arises from the fact that a image contains many similar yet non-local patches, was widely used for image denoising. Some well-known NSS based method include BM3D \cite{dabov2007image} and NLM \cite{buades2005non}. Other methods, such as Gaussian Mixture Model (GMM) \cite{zhu2016noise, zhao2014robust, meng2013robust} and dictionary 
learning \cite{dong2011sparsity} had also been applied to model the noise distribution. The performance of these model-based methods is depend on the accuracy of their prior assumptions. Once they encounter a dataset with an unknown noise distribution, their performance is limited.

Learning-based methods have achieved state-of-the-art denoising results through deep convolutional neural networks \cite{batson2019noise2self, krull2019probabilistic, yue2019variational, brooks2019unprocessing}. In \cite{zhang2017beyond}, Zhang \emph{et al.} propose DnCNN that employs a very deep CNN to remove different levels of Gaussian noise. In \cite{zhang2018ffdnet}, Zhang \emph{et al.} exploit a tunable noise level map as the input to help the performance of a non-blind denoising network. In \cite{guo2019toward}, Guo \emph{et al.} develop CBDNet that consists of one noise level estimator and one non-blind denoising network to handle diverse noise. The training of these CNNs requires a large amount of paired data, which makes it difficult for them to be applied in practical tasks. Later, Lehtinen \emph{et al.} \cite{lehtinen2018noise2noise} demonstrate that clean targets can be absent. Their Noise2Noise training scheme allows the network to learn the mapping between two instances of the noisy image containing the same signal. However, the acquisition of two noisy realizations of the same image content is often impractical. More recently, Krull \emph{et al.} \cite{krull2019noise2void} extend Noise2Noise and propose a self-supervised method that can train a denoiser using only individual noisy images, namely Noise2Void (N2V). The effectiveness of N2V is based on a strong assumption that pixels values of the noise is zero mean and independent between pixels. Krull \emph{et al.} further points out that N2V fails to remove noise that violates their assumptions, such as structured noise. In this respect, our N2B is robust to many types of noise as long as there are some unpaired noisy/clean images.
\begin{figure}
	\centering
	\subfigure[Initial stage.]{\includegraphics[width=6in]{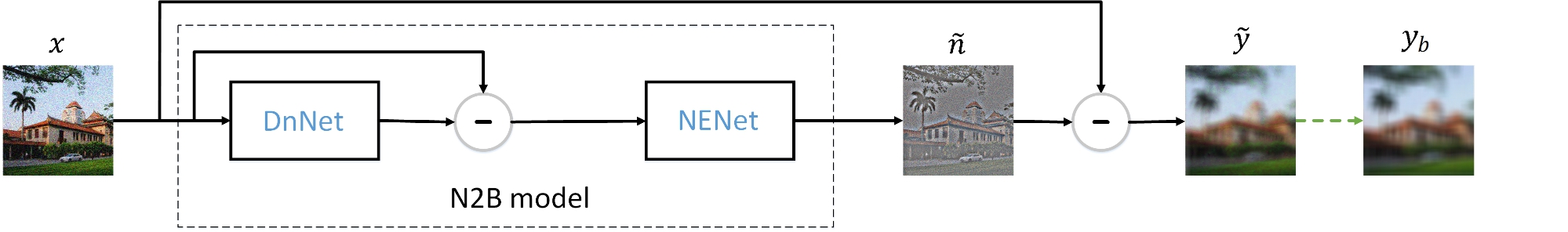}\label{fig.frame_a}}
	\subfigure[Convergence stage.]{\includegraphics[width=7in]{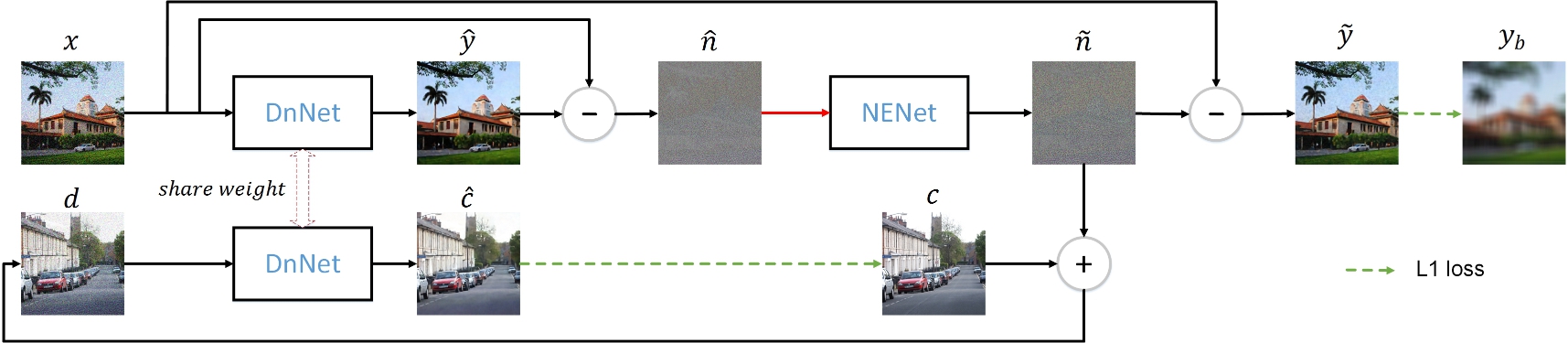}\label{fig.frame_b}}
	\caption{The framework of our Noise2Blur for training denoising model. (top) First, we use a image filter to create a blurred label for supervision. The result is a noise-free but severely blurred image. (bottom) We add noise map extracted by the N2B model to a random clean image to generate a new ``noise-to-clean'' objective. The N2B model is then divided into a denoising network (DnNet) and a noise extraction network (NENet). We use the DnNet to learn ``noise-to-clean'', while NENet is supervised by ``noise-to-blur''. These two networks cyclically use each other's knowledge to promote their own learning. The red line ``\textcolor{red}{$\rightarrow$}'' means that it can only propagate forward but the gradient cannot be backpropagated.
	}
	\label{fig_frame}
\end{figure}


\section{Proposed method}
For denoising problems, the image degradation model is typically assumed as $x=y+n$ \cite{gu2019brief, krull2019noise2void}. Image denoising attempts to restore the noise-free image $y$ from its noisy observation $x$ by removing the noise $n$. This inverse problem is tricky because the statistical information of noise is often unknown (``blind'') in real-world applications. Here, we propose Noise2Blur (N2B), a training framework for blind denoising. Unlike previous denoising methods \cite{guo2019toward, zhang2018ffdnet, chen2018image}, N2B training requires neither a large amount of pre-collected paired data nor an estimate of the noise model. The only prerequisite is some unpaired noisy and clean images, which is easy to obtain. The basic idea of our approach is to extract noise directly from noisy data to construct a paired training data for denoising. This is done by two networks: one for denoising and the other for noise extraction. To begin training, we use blurred labels to generate the necessary supervision, which can be obtained by filtering a noisy input. The training process of N2B model can be divided into two stages:

\begin{itemize}
	\item Initial stage: We use the blurred labels to allow the N2B model to \textit{roughly} remove noise (``noise-to-blur'').
	\item Convergence stage: The noise information output from the N2B model is added to a random clean image to generate a new ``noise-to-clean'' supervision. Combining clean and blurred objectives, the N2B model is divided into two parts, \emph{i.e.} a denoising network and a noise extraction network. By a simple gradient interruption operation, these two networks separately serve their respective objectives, a denoising network for ``noise-to-clean'' and a noise extraction network for ``noise-to-blur''.
\end{itemize}
In the following subsections, we will introduce the details of the N2B implementation.

\subsection{Initial stage}

Assuming that only some unpaired noisy and clean images are available, we cannot use these data directly to learn the noise removal function. To get around this problem, we use some image prior knowledge to generate extra information for learning. Specifically, we know that noise can be easily removed by some general image filtering techniques such as mean filter, Gaussian filter, or bilateral filter. Although many high frequency image details are also filtered out, the simple filtering techniques can indeed introduce new knowledge into the denoising process. For the N2B training (Figure \ref{fig.frame_a}), the noisy observation $x$ is first transformed into a noise-free but blurred label $y_b$ by an image filter,
\begin{equation}
\label{eq.yb}
y_b=x- n_b,
\end{equation}

\begin{equation}
\label{eq.yb2}
n_b=n+\epsilon_b,
\end{equation}
where $n_b$ is a noise component lost by filtering, $\epsilon_b$ denotes some image details from $x$. $y_b$ and $x$ can be paired to provide supervision. Given the noisy input $x$, the direct output of the N2B model is a noise map $\tilde{n}$, 
\begin{equation}
\label{eq.n}
\tilde{n}=S(x; \theta_S),
\end{equation}
where $S(\cdot)$ denotes the mapping function of the N2B model, $\theta_S$ represents the weight parameters. The corresponding noise removal result $\tilde{y}$ is obtained by
\begin{equation}
\tilde{y}=x-\tilde{n}.
\end{equation}

In this initial stage, the N2B training is guided by the following ``noise-to-blur'' objective,
\begin{equation}
\label{eq.n2b}
L_{n2b}=\frac{1}{M}\sum_{i=1}^{M}\Vert \tilde{y}^i-y_b^i \Vert,
\end{equation}
where $M$ is the number of training noisy images, and we adopt $l_1$ loss \cite{zhao2016loss} for training. By a simple transformation, Eq. (\ref{eq.n2b}) is equivalent to
\begin{equation}
\label{eq.n2b2}
L_{n2b}=\frac{1}{M}\sum_{i=1}^{M}\Vert \tilde{n}^i-n_b^i \Vert.
\end{equation}

\begin{figure}[t]
	\centering
	\subfigure[]{\includegraphics[width=1in]{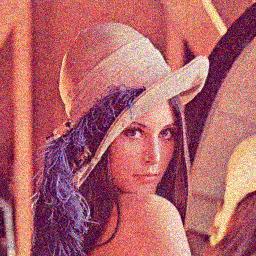}}
	\subfigure[]{\includegraphics[width=1in]{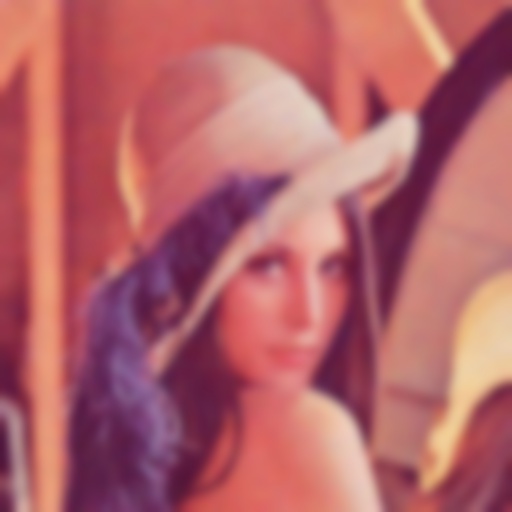}}
	\subfigure[]{\includegraphics[width=1in]{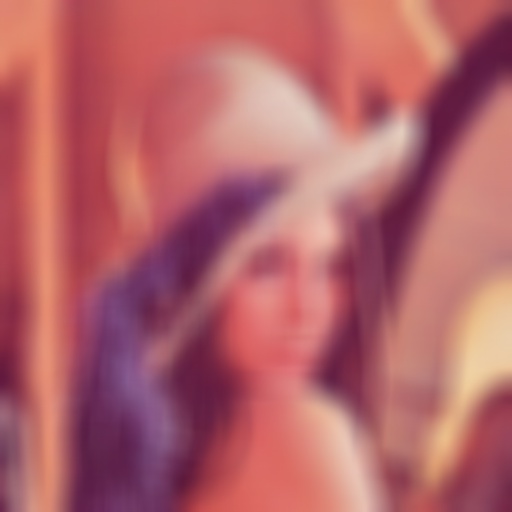}}
	\subfigure[]{\includegraphics[width=1in]{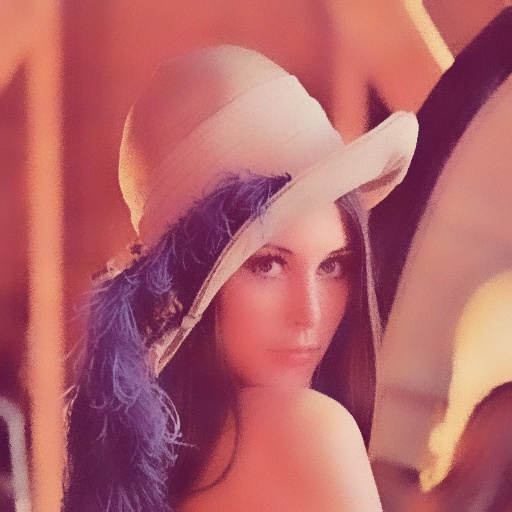}}
	\subfigure[]{\includegraphics[width=1in]{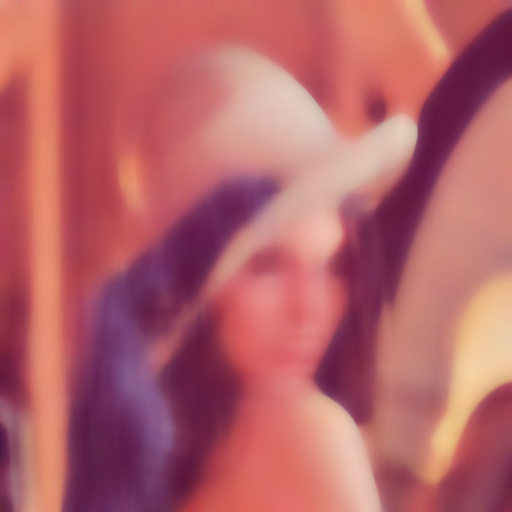}}
	\subfigure[]{\includegraphics[width=1in]{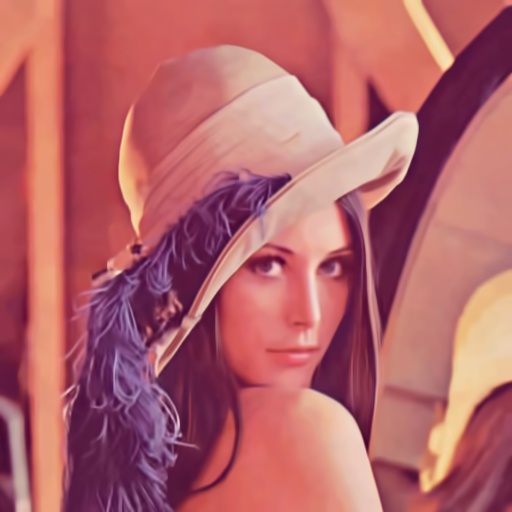}}

	\caption{Blurred images generated using different methods. (a) Gaussian noise, $\sigma=25$. (b) Gaussian filter, kernel size=31. (c) Mean filter, kernel size=31. (d) Bilateral filter. (e) Median filter, kernel size=31. (f) BM3D.}
	\label{fig_lenna}
\end{figure}

Guided by the image filter, the N2B model learns to remove noise along with much image detail. Although the results are not satisfactory, the N2B model is indeed learning to remove noise. This learning process can be seen as finding a good initial state for the model. On the other hand, our N2B model has no requirement for the quality of blurred labels. Therefore, any filter can be adopted as long as it can filter out noise cleanly (see Figure \ref{fig_lenna}).


\subsection{Convergence stage}
Under the supervision of Eq. (\ref{eq.n2b2}), the N2B model learns to remove noise, but is not able to restore the lost image details. To enhance image quality while removing noise, some other supervision is therefore needed. 

We observe that the N2B model can be regarded as a noise extractor. The noise map $\tilde{n}$ Eq. (\ref{eq.n}) output by the N2B model can be further represented as
\begin{equation}
\label{eq.ntilde}
\tilde{n}=n + \tilde{\epsilon},
\end{equation}
where $\tilde{\epsilon}$ denotes some image details from the noisy input $x$. Based on Eqs. (\ref{eq.yb2}) and (\ref{eq.n2b2}), $\tilde{\epsilon}$ is expected to be equal to $\epsilon_b$. Note that $\tilde{n}$ contains all the noise information $n$ of the input noisy image $x$, and we can use it to generate extra knowledge. In this respect, we need some pre-collected clean images. For each noisy input $x$, the noise map $\tilde{n}$ is added to a random clean image $c$ to generate a new noisy image (see Figure \ref{fig.frame_b})
\begin{equation}
\label{eq.caddn}
d=c+\tilde{n}=c+n+\tilde{\epsilon}.
\end{equation}

\begin{algorithm*}[t] 
	\caption{Convergence stage for N2B training.} 
	\label{algorithm1} 
	\begin{algorithmic}[1] 
		\REQUIRE Unpaired noisy and clean datasets. Blurred labels obtained by predefined filtering of the noisy images.
		
		\FOR{$t=1$ \textbf{to} $T$} 
		
		\STATE Sample minibatch of $m$ noisy images $\{x^1,...,x^m\}$ from the noisy dataset.
		
		\STATE Input $\{x^1,...,x^m\}$ to DnNet, and generate $\{\hat{n}^1,...,\hat{n}^m\}$ with Eq. (\ref{eq.nhat}).
		
		\STATE Input $\{\hat{n}^1,...,\hat{n}^m\}$ to NENet, and generate $\{\tilde{n}^1,...,\tilde{n}^m\}$ with Eq. (\ref{eq.ntil}).
		
		\STATE Sample minibatch of $m$ clean images $\{c^1,...,c^m\}$ from the clean dataset.
		
		\STATE Add $\{\tilde{n}^1,...,\tilde{n}^m\}$ to $\{c^1,...,c^m\}$ to get new noisy images $\{d^1=c^1+\tilde{n}^1,...,d^m=c^m+\tilde{n}^m\}$.
	    
	    \STATE Input $\{d^1=c^1+\tilde{n}^1,...,d^m=c^m+\tilde{n}^m\}$ to DnNet, and generate $\{\hat{c}^1,...,\hat{c}^m\}$ with Eq. (\ref{eq.c_hat}).
	    	
		\STATE Disconnect the gradient connection between DnNet and NENet.
		
		\STATE Update the DnNet and NENet with Eqs. (\ref{eq.n2c}) and (\ref{eq.n2b3}).

		\ENDFOR
		
	\end{algorithmic}
\end{algorithm*}

In this way, we get a new noisy/clean image pair. At each epoch, $x$ has a new counterpart $d$, they have the same noise component $n$, but their image content is different. Also, the label for $x$ is blurred, while $d$ has a clean label $c$. We then use the new image pair $d$ and $c$ to provide a ``noise-to-clean'' objective for denoising. However, “noise-to-clean” and “noise-to-blur” are contradictory; they cannot work in tandem for the same model. To solve this problem, we divide the N2B model into two parts, \emph{i.e.} a denoising network (DnNet) and a noise extraction network (NENet), to serve each objective. 

DnNet is guided to restore image details while removing noise by the loss,
\begin{equation}
\label{eq.n2c}
L_{n2c}=\frac{1}{M}\sum_{i=1}^{M}\Vert \hat{c}^i-c^i \Vert,
\end{equation}
\begin{equation}
\label{eq.c_hat}
\hat{c}=F(d; \theta_F),
\end{equation}
where $F(\cdot)$ denotes the mapping function of the DnNet, $\theta_F$ represents the weight parameters. When inputting a noisy image, the desired output of DnNet should be a noise-free and high-quality image. However, the performance of DnNet depends on its training data, \emph{i.e.} $d$ and $c$. According to Eq. (\ref{eq.caddn}), we hope the $\tilde{\epsilon}$ in $\tilde{n}$ to be as weak as possible, so that $d$ will be a more realistic noisy image. This needs to be achieved by the NENet.

NENet is expected to output a noise map, so we call it noise extraction network. Since only blurred labels are available, NENet is supervised by the “n2b" objective Eq. (\ref{eq.n2b2}). The input to NENet is a noise map $\hat{n}$, which is generated by DnNet,
\begin{equation}
\label{eq.nhat}
\hat{n}=x-F(x; \theta_F)=n+\hat{\epsilon},
\end{equation}
where $\hat{\epsilon}$ denotes some image details from the noisy input $x$.
Based on Eqs. (\ref{eq.n2b2}) and (\ref{eq.ntilde}), the output of NENet is the noise map $\tilde{n}$,
\begin{equation}
\label{eq.ntil}
\tilde{n}=H(\hat{n}; \theta_H)=n+\tilde{\epsilon},
\end{equation}
where $H(\cdot)$ denotes the mapping function of NENet, $\theta_H$ represents the weight parameters. Based on Eqs. (\ref{eq.yb2}), (\ref{eq.nhat}) and (\ref{eq.ntil}), the “n2b" loss Eq. (\ref{eq.n2b2}) can be rewritten as

\begin{equation}
\label{eq.n2b3}
\begin{aligned}
L_{n2b}&=\frac{1}{M}\sum_{i=1}^{M}\Vert \tilde{n}^i-n_b^i \Vert \\
&=\frac{1}{M}\sum_{i=1}^{M}\Vert H(\hat{n}^i; \theta_H)-n_b^i \Vert \\
&=\frac{1}{M}\sum_{i=1}^{M}\Vert H(n^i+\hat{\epsilon}^i; \theta_H)-(n^i+\epsilon_b^i) \Vert.
\end{aligned}
\end{equation}

\begin{figure}[t]
	\centering
	\subfigure[$x$]{\includegraphics[width=0.8in]{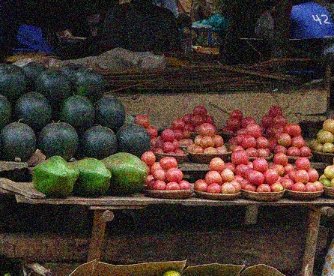}}
	\subfigure[$n_b$]{\includegraphics[width=0.8in]{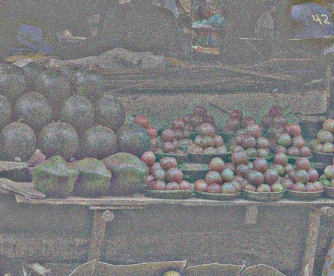}}
	\subfigure[$\hat{n}$]{\includegraphics[width=0.8in]{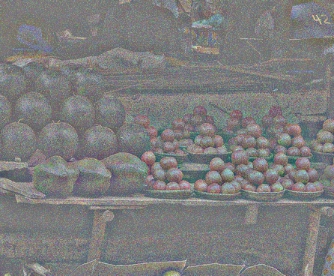}}
	\subfigure[$\hat{n}^{100}$]{\includegraphics[width=0.8in]{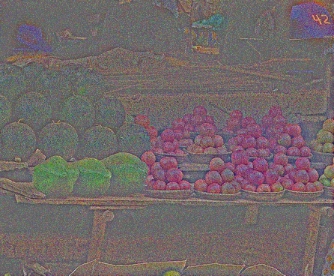}}
	\subfigure[$\hat{n}^{300}$]{\includegraphics[width=0.8in]{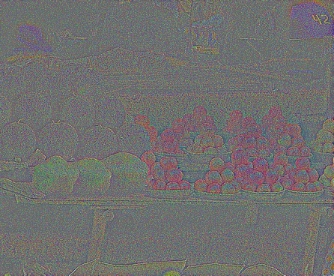}}
	\subfigure[$\hat{n}^{1k}$]{\includegraphics[width=0.8in]{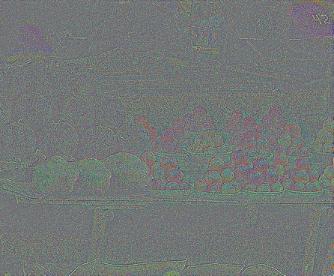}}
	\subfigure[$\hat{n}^{15k}$]{\includegraphics[width=0.8in]{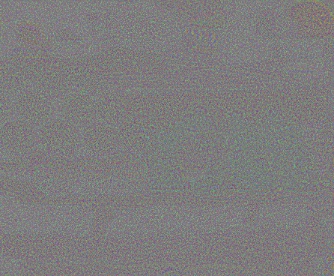}}

	\caption{Illustrations of the effect of gradient interruption. (a)-(b): Noisy image $x$ and the noise map $n_b=n+\epsilon_b$ filtered by a bilateral filter. (c): $\hat{n}=n+\hat{\epsilon}$. Without gradient interruption, DnNet has to make a trade-off between $L_{n2c}$ and $L_{n2b}$, which causes the noise map $\hat{n}$ generated by DnNet to contain some image details $\hat{\epsilon}$. (d)-(g): Using gradient interruption, DnNet learns to preserve image details while removing noise, so the image detail $\hat{\epsilon}$ in $\hat{n}$ gradually decays. As training progresses, the $\hat{\epsilon}$ will be too weak to support NENet to meet ``n2b'' objective. The superscript of $\hat{n}$ represents the number of iterations in the convergence stage.}
	\label{fig_gi}
\end{figure}

In Eq. (\ref{eq.n2b3}), the input ($n+\hat{\epsilon}$) and target ($n+\epsilon_b$) for NENet both contain the same component $n$. To reduce the value of $L_{n2b}$, NENet learns to transform $\hat{\epsilon}$ into $\epsilon_b$ while maintaining $n$. However, $\hat{\epsilon}$ is generated by DnNet Eq. (\ref{eq.nhat}), and $L_{n2c}$ Eq. (\ref{eq.n2c}) encourages DnNet to preserve image details when denoising, which is equivalent to reducing $\Vert \hat{\epsilon} \Vert$ to 0. Note that, if $\Vert \hat{\epsilon} \Vert$ converges to 0, “n2b" Eq. (\ref{eq.n2b3}) will be an impossible task for NENet. Intuitively, if there is a gradient connection between the two networks (\textit{i.e.} the gradient calculated by $L_{n2b}$ Eq. (\ref{eq.n2b3}) can be propagated back to DnNet), the  $L_{n2c}$ and $L_{n2b}$ will be adversarial. DnNet has to make a trade-off between $L_{n2c}$ and $L_{n2b}$, which results in $\Vert \hat{\epsilon} \Vert$ should be greater than 0. Only in this case can NENet reduce $L_{n2b}$ by transforming $\hat{\epsilon}$ into $\epsilon_b$ (see Figure \ref{fig_gi}(a)-(c)). When $L_{n2b}$ converges to a low value, the noise map $\tilde{n}$ Eq. (\ref{eq.ntil}) output by NENet will contain not only the noise component $n$ but also many image details $\tilde{\epsilon}$ from noisy image $x$. However, we hope that the noise map $\tilde{n}$ extracted by NENet is mainly composed of the noise $n$, \textit{i.e.}
\begin{equation}
\label{eq.ntild3}
\tilde{n}=H(\hat{n}; \theta_H)=H(n+\hat{\epsilon}; \theta_H)\approx n,
\end{equation}
so that we can obtain realistic noisy image $d$ to train DnNet. To this end, we apply a simple gradient interruption to training. In this convergence stage, $\hat{n}$ can be propagated forward to NENet, but the gradient calculated by the $L_{n2b}$ can no longer be propagated back to DnNet. Due to gradient interruption, $L_{n2b}$ cannot inform DnNet to generate the noise map $\hat{n}$ with the specific image details $\hat{\epsilon}$. In addition, DnNet can focus on $L_{n2c}$, so it learns to preserve image details when removing noise. This means that the image detail $\hat{\epsilon}$ in the noise map $\hat{n}$ generated by DnNet will gradually decay (see Figure \ref{fig_gi} (d)-(g)). Therefore, the information contained in the input noise map $\hat{n}$ will not be sufficient to support NENet to meet the ``n2b'' objective Eq. (\ref{eq.n2b3}). In other words, as training progresses, it is difficult for NENet to transform the continuously decaying $\hat{\epsilon}$ into $\epsilon_b$. According to Eq. (\ref{eq.n2b3}), NENet can still maintain $n$. To avoid incorrect results, the output of NENet tends to be similar to its input (\emph{i.e.} $\tilde{n}\approx \hat{n}$ and $\tilde{y}\approx \hat{y}$), which also means that the image details $\tilde{\epsilon}$ in $\tilde{n}$ will gradually decay. After convergence, DnNet has the ability to restore high-quality images (\textit{i.e.} $\hat{\epsilon}\approx 0$) while removing noise. Since $\hat{\epsilon}\approx 0$, NENet cannot meet the ``n2b'' objective, it can only learn to extract noise, \textit{i.e.}


\begin{equation}
\label{eq.ntild4}
\tilde{n}=H(n+\hat{\epsilon}; \theta_H)\approx H(n+0; \theta_H)\approx n.
\end{equation}

Eq. (\ref{eq.ntild4}) implies that the $L_{n2b}$ will converge to 

\begin{equation}
\label{eq.n2b5}
\begin{aligned}
L_{n2b}&=\frac{1}{M}\sum_{i=1}^{M}\Vert H(n^i+\hat{\epsilon}^i; \theta_H)-(n^i+\epsilon_b^i) \Vert\\
&\approx \frac{1}{M}\sum_{i=1}^{M}\Vert n^i-(n^i+\epsilon_b^i) \Vert \approx \frac{1}{M}\sum_{i=1}^{M}\Vert \epsilon_b^i \Vert.
\end{aligned}
\end{equation}


\begin{figure}[t]
	\centering
	\includegraphics[width=4in]{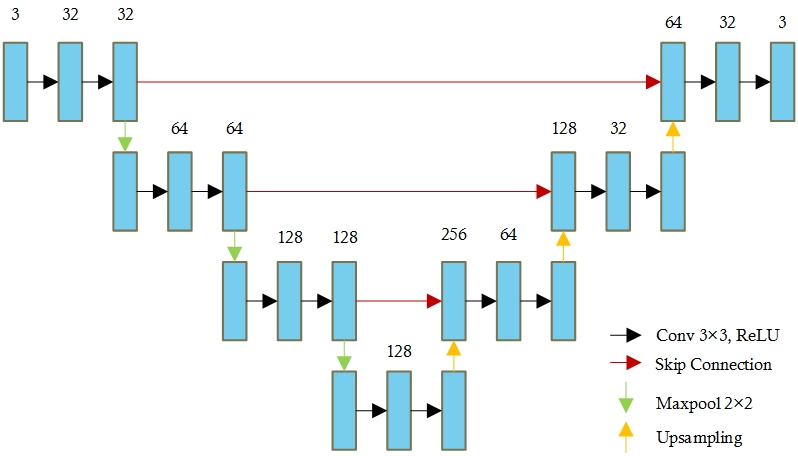}

	\caption{The U-Net architecture of our denoising network.}
	\label{unet}
	
\end{figure}

The procedure of convergence stage is shown in Algorithm \ref{algorithm1}. We note that, during testing, we remove NENet, and retain only DnNet to obtain the final denoising result. 

\subsection{Parameters settings and training details}
N2B can be applied using any network architecture. We choose a simple U-Net as the denosing network for our experiments (see Figure \ref{unet}), while the noise extraction network is composed of three $3 \times 3$ convolution layers with the number of feature map is 32. We use PyTorch and Adam \cite{kingma2014adam} with a mini-batch size of 16 to train our model. We randomly select $128\times 128$ image patches from training datasets as inputs. In training, the initial stage makes up $5\%$ of the training time, while the remaining $95\%$ is the convergence stage.

\section{Experiments}
We evaluate the performance of Noise2Blur on several denoising tasks in this section. Our Noise2Blur results are compared to the results of several model-based and CNN-based methods. Moreover, we show that Noise2Blur is a general technique that can be applied to diverse image restoration problems, such as text removal.

\subsection{Synthetic noises}
To synthesize noisy images, we use $4744$ natural images provided by \cite{ma2017waterloo} as the training set. Noisy versions of all images are generated by adding random noise sampled from different distribution. In addition, we collected another $4000$ clean natural images from the Internet as a clean set for N2B training, we call it $Clean4000$. Note that the images in $Clean4000$ are different from the 4744 images \cite{ma2017waterloo}. To learn to remove synthetic noise, our N2B model is trained for $1000$ epochs, and the first $50$ epochs being the initial stage. We compare our method with several state-of-the-art denoising methods: BM3D \cite{dabov2007image}, WNNM \cite{gu2014weighted}, DnCNN \cite{zhang2017beyond}, Noise2Noise (N2N) \cite{lehtinen2018noise2noise}, Noise2Void (N2V) \cite{krull2019noise2void} and a U-Net trained in a standard supervised manner using paired noisy/clean images. In particular, N2N, U-Net and our N2B use the same network architecture (see Figure \ref{unet}). For BM3D, we set its hyperparameter $\sigma$ to 25 when testing Gaussian noise($\sigma=25$), while in other experiments, BM3D kept its default setting (\textit{i.e.} $\sigma=50$). Our test dataset is the commonly used BSD300 \cite{martin2001database}.

\begin{figure*}[t]
	\centering
	
	\subfigure[\scriptsize {Clean $|$ SSIM, PSNR}]{\includegraphics[width=1.5in]{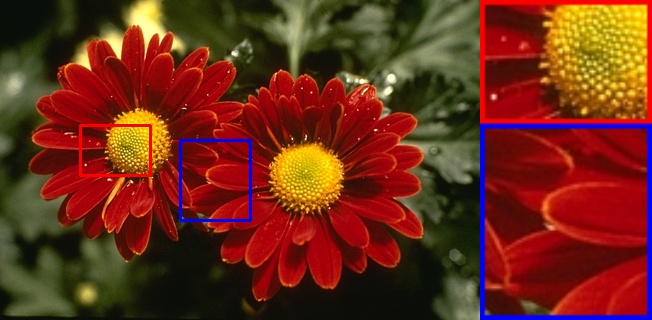}}
	\subfigure[\scriptsize {Input $|$ 0.503, 24.58}]{\includegraphics[width=1.5in]{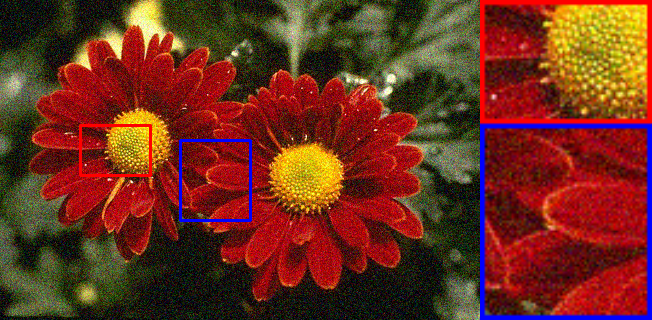}}
	\subfigure[\scriptsize {BM3D $|$ 0.904, 31.71}]{\includegraphics[width=1.5in]{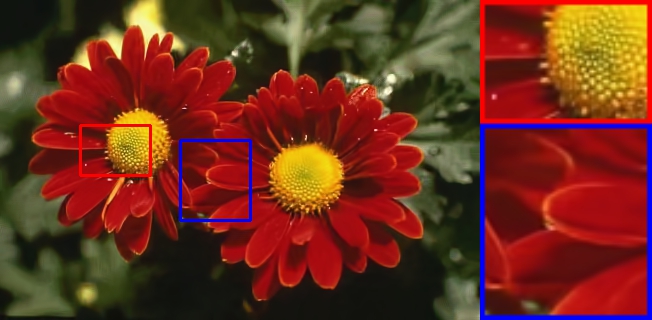}}
	\subfigure[\scriptsize {DnCNN $|$ \textbf{0.922}, \textbf{33.58}}]{\includegraphics[width=1.5in]{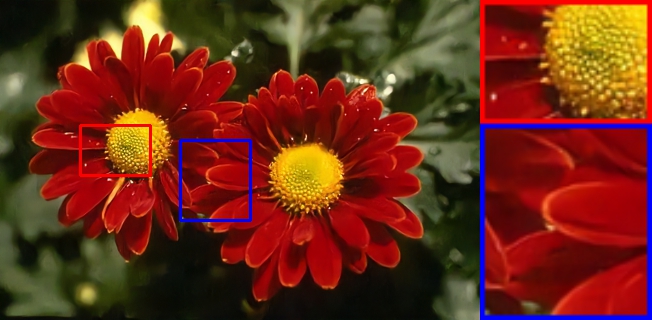}}
	\\
	\subfigure[\scriptsize {N2V $|$ 0.835, 29.61}]{\includegraphics[width=1.5in]{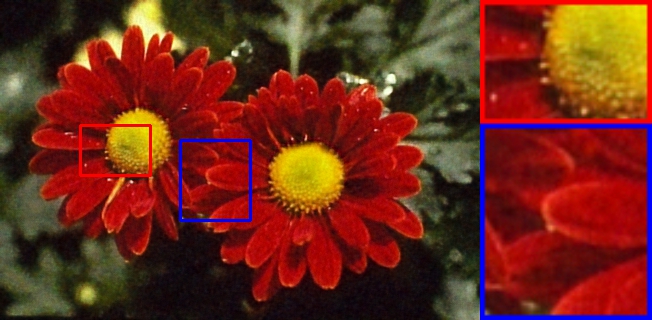}}
	\subfigure[\scriptsize {U-Net $|$ 0.920, 32.95}]{\includegraphics[width=1.5in]{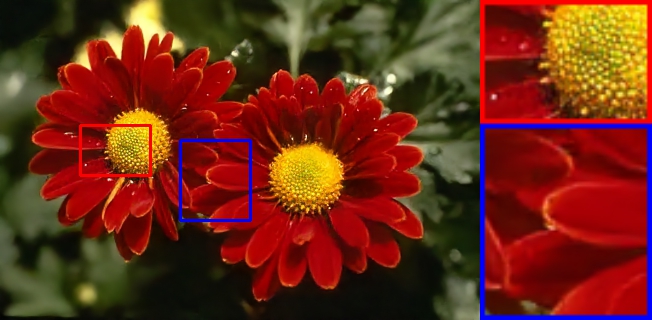}}
	\subfigure[\scriptsize {N2B$_{one}$ $|$ 0.865,31.63}]{\includegraphics[width=1.5in]{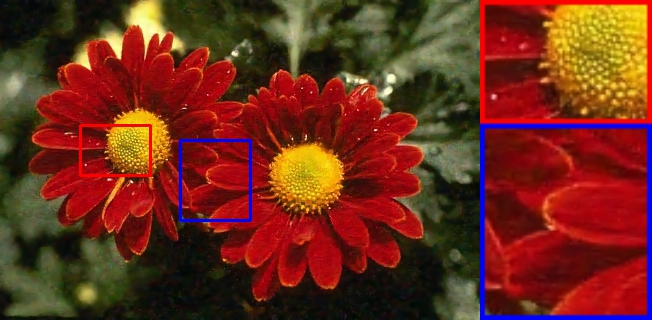}}
	\subfigure[\scriptsize {N2B $|$ 0.905, 32.22}]{\includegraphics[width=1.5in]{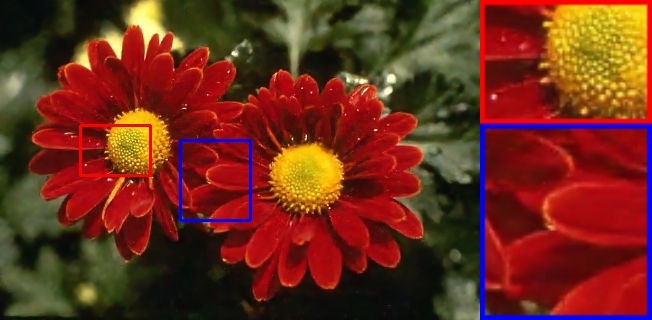}}
	
	\caption{One visual comparisons for Gaussian denoising, $\sigma=25$.}
	\label{fig_gaus}
\end{figure*} 

\begin{table*}[t]
	\caption{PSNR results (dB) from BSD300 dataset for Gaussian, Speckle and Salt \& pepper noise. (N2B$_{one}$ is trained using only one $512\times512$ noisy image, with Gausssian noise $\sigma=25$, Speckle noise $v=0.1$ or Salt \& pepper noise $p=0.15$, respectively.)}
	\centering
	\begin{tabular}{|c|c|c|c|c|c|c|c|c|c|}
		\hline
		&Test noise level& BM3D & WNNM&DnCNN&N2N&N2V&U-Net&N2B$_{one}$&N2B\\
		\hline
		\multirow{2}*{Gaussian }&$\sigma=25$&30.90&29.96&\textbf{31.54}&31.32&28.14&31.45&30.11&30.87\\
		\cline{2-10}
		&$\sigma \in [0, 50]$&27.89&31.16&\textbf{33.21}&32.82&28.11&33.14&30.01&31.98\\
		\hline
		\multirow{2}*{Speckle }&$v=0.1$&26.64&25.13&\textbf{31.21}&31.12&27.10&31.18&28.59&29.50\\
		\cline{2-10}
		&$v \in [0, 0.2]$&26.70&25.39&\textbf{31.59}&31.50&27.56&31.55&28.05&29.81\\
		\hline
		\multirow{2}*{Salt \& pepper }&$p=0.15$&24.68&36.16&\textbf{44.96}&44.37&26.72&44.10&32.09&36.22\\
		\cline{2-10}
		&$p \in [0, 0.3]$&24.37&36.39&\textbf{45.53}&44.85&27.27&44.65&31.61&36.53\\
		\hline
		
	\end{tabular}
	\label{table1}
\end{table*}


\begin{figure}[t]
	\centering
	\includegraphics[width=.3\columnwidth]{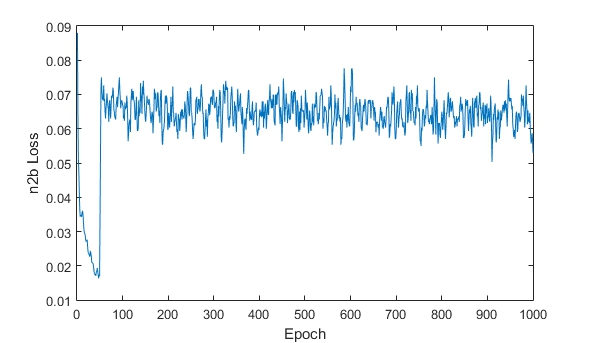}
	\includegraphics[width=.3\columnwidth]{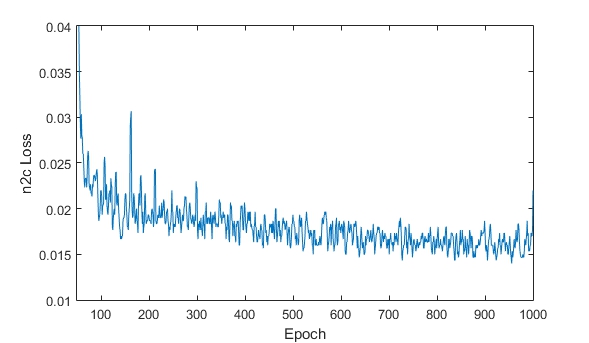}
	\caption{The loss curves for Gaussian denoising. Left: ``n2b'' loss (\ref{eq.n2b3}). Right: ``n2c'' loss (\ref{eq.n2c}).}
	\label{n2b_curve}
\end{figure}

\textit{\textbf{Gaussian noise.}}  We first conduct experiments on a blind Gaussian denoising. Each training example is degraded by Gaussian noise with a random standard deviation $\sigma \in [0,50]$. For the N2B training, we use the bilateral filter to generate the blurred labels. Table \ref{table1} shows the output image quality for the various methods. During testing, the test noisy images are generated with two different types of noise levels, \emph{i.e.} a fixed $\sigma=25$ and a variable $\sigma \in [0, 50]$.  As can be seen, our N2B significantly outperform the self-supervised method (N2V) and the prior-based methods (BM3D and WNNM). Although our method is currently unable to achieve higher numerical metrics than noise2noise and fully-supervised U-Net (``noise2clean''), the results are satisfactory. In many real-world cases, the noise2noise training and fully-supervised training are often impractical, due to the lack of paired noisy/noisy or noisy/clean images. Our method can still achieve decent results in the absence of paired 
data, which makes it a flexible and promising option in practical application. On the other hand, we find that N2B training does not require a large amount of noisy images. Even if we only have one $512\times 512$ Lena image (see Figure \ref{fig_lenna}) with Gaussian noise $\sigma=25$ for training, the N2B model (\emph{N2B$_{one}$}) still has strong generalization. This is because our N2B training can synthesize many new noisy images online, which can be regarded as data augmentation. Therefore, it greatly relaxes the need of pre-collected noisy images for training CNNs. We show the visual results in Figure \ref{fig_gaus}. Our method achieves promising results in removing noise and enhancing image quality.

In addition, we show the curves of the ``n2b'' loss Eq. (\ref{eq.n2b3}) and ``n2c'' loss Eq. (\ref{eq.n2c}) in training to demonstrate the idea of this paper (see Figure \ref{n2b_curve}). As can be seen, the ``n2c'' loss gradually decreases as the training progresses. In contrast, the ``n2b'' loss can hardly descend in the convergence stage ($50-1000$ epochs) and converges to a high value. This is because the information in the input $\hat{n}$ of the NENet is not enough to support the NENet to meet the ``n2b'' objective Eq. (\ref{eq.n2b3}). On the hand, we also try to compare the denoising results generated by DnNet and NENet (\emph{i.e.} $\hat{y}$ and $\tilde{y}$). The NENet achieves an average quality of $29.22dB$ with $\sigma=25$, close to $30.86dB$ of DnNet. This means that NENet fails to blur the image, all NENet can do is maintain the noise map (\emph{i.e.} $\tilde{n}\approx \hat{n}$ and $\tilde{y}\approx \hat{y}$). 

\begin{figure*}[t]
	\centering
	
	\subfigure[\scriptsize {Clean $|$ SSIM, PSNR}]{\includegraphics[width=1.5in]{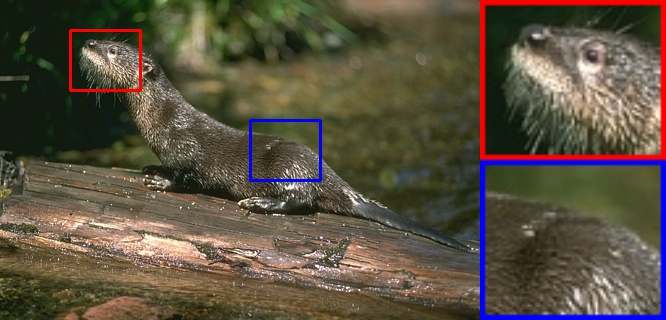}}
	\subfigure[\scriptsize {Input $|$ 0.535, 22.47}]{\includegraphics[width=1.5in]{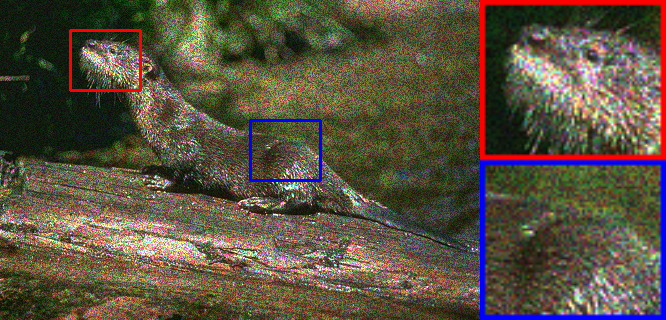}}
	\subfigure[\scriptsize {BM3D $|$ 0.787, 26,39}]{\includegraphics[width=1.5in]{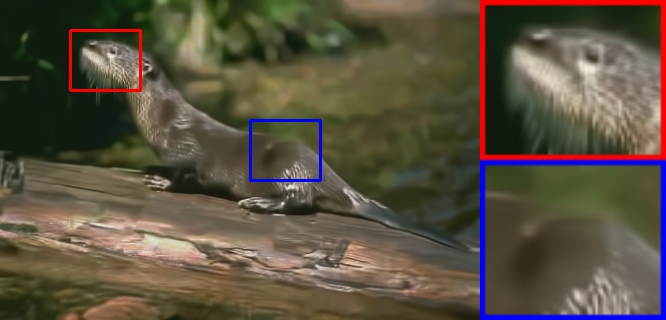}}
	\subfigure[\scriptsize {DnCNN $|$ 0.920, \textbf{30.36}}]{\includegraphics[width=1.5in]{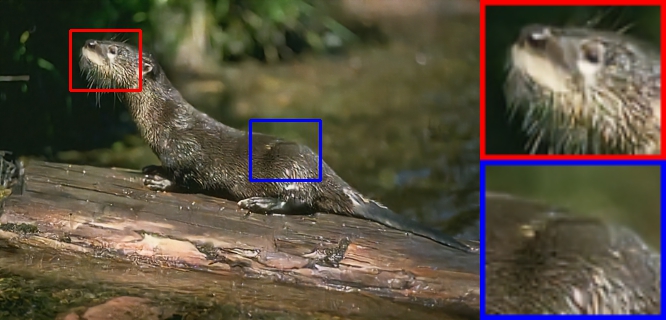}}
	\\
	\subfigure[\scriptsize {N2V $|$ 0.819, 27.20}]{\includegraphics[width=1.5in]{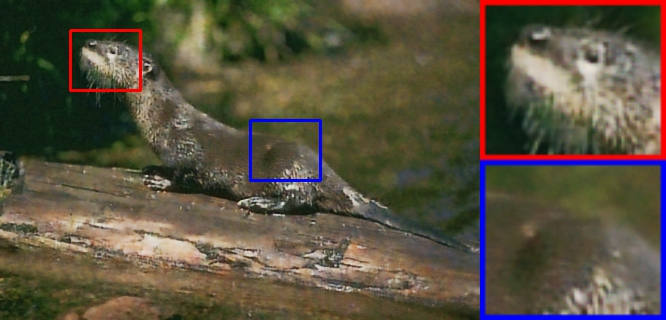}}
	\subfigure[\scriptsize {U-Net $|$ \textbf{0.925}, 30.34}]{\includegraphics[width=1.5in]{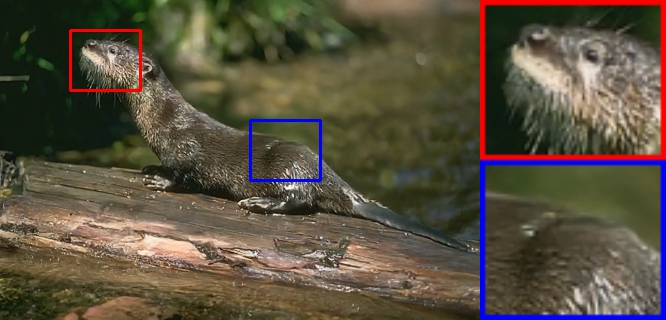}}
	\subfigure[\scriptsize {N2B$_{one}$ $|$ 0.875, 28.67}]{\includegraphics[width=1.5in]{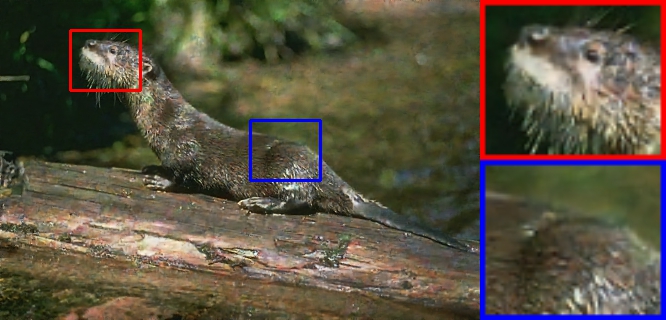}}
	\subfigure[\scriptsize {N2B $|$ 0.895, 29.10}]{\includegraphics[width=1.5in]{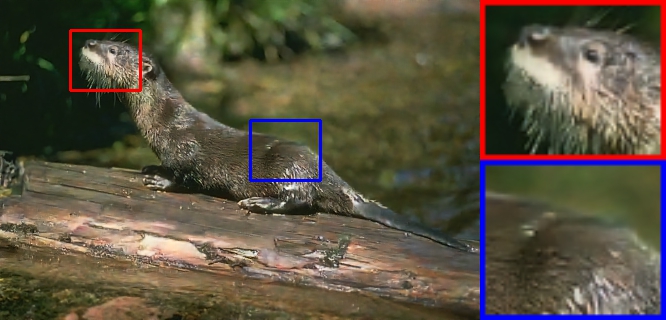}}

	\caption{One visual comparisons for Speckle denoising, $v=0.1$.}
	\label{fig_speckle}
\end{figure*} 

\textit{\textbf{Speckle noise.}} Speckle is a granular interference that inherently degrades the quality of an active radar \cite{argenti2013tutorial}, synthetic aperture radar \cite{lee2014polarimetric}, and ultrasound images \cite{wu2013evaluation}. It is typically known as a multiplicative noise to the image $y$, which can be modeled using the equation $x=y+n\cdot y$. In this equation, $n$ is uniformly distributed random noise with mean 0 and variance $v$. We synthesize noisy images for training by varying the noise variance $v\in [0, 0.2]$. Following previous publications \cite{chierchia2017sar, yang2019sar}, we use a coupled log and exp transforms to simplify this denoising problem. For our N2B model, a mean filter with the kernel size 35 is adopted to generate blurred labels. The results are shown in Figure \ref{fig_speckle} and Table \ref{table1}. As observed, our N2B consistently shows promising performance. 

\begin{figure*}[t]
	\centering
	
	\subfigure[\scriptsize {Clean $|$ SSIM, PSNR}]{\includegraphics[width=1.5in]{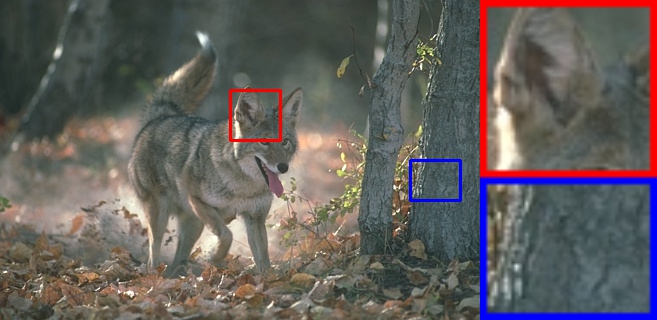}}
	\subfigure[\scriptsize {Input $|$ 0.224, 17.15}]{\includegraphics[width=1.5in]{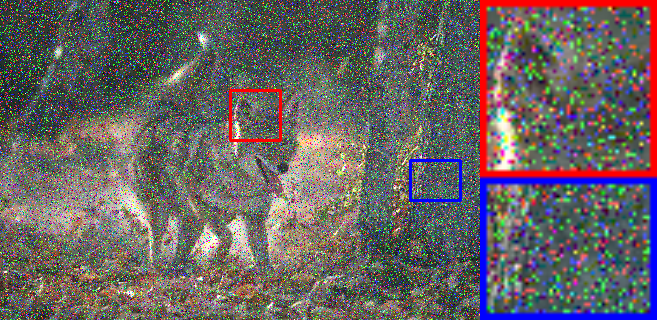}}
	\subfigure[\scriptsize {BM3D $|$ 0.722, 26.08}]{\includegraphics[width=1.5in]{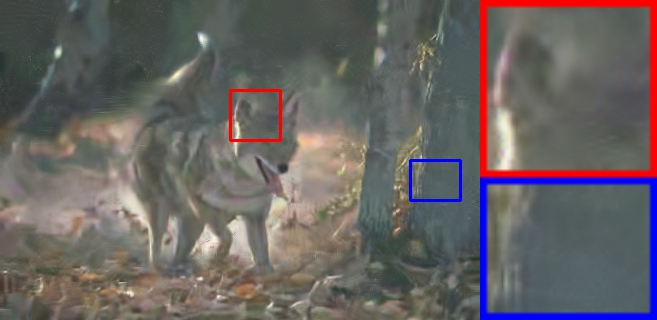}}
	\subfigure[\scriptsize {DnCNN $|$ \textbf{0.986}, \textbf{45.53}}]{\includegraphics[width=1.5in]{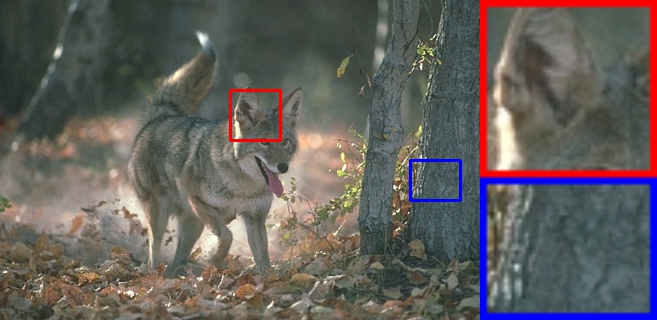}}
	\\
	\subfigure[\scriptsize {N2V $|$ 0.864, 28.67}]{\includegraphics[width=1.5in]{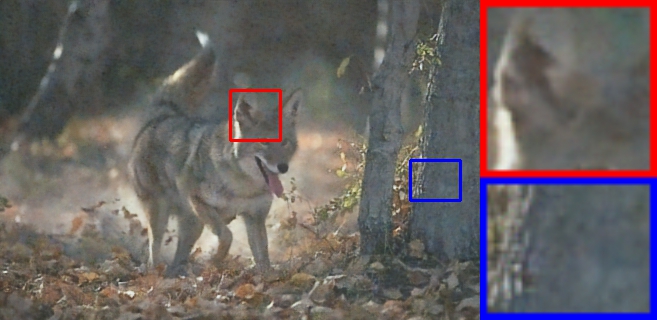}}
	\subfigure[\scriptsize {U-Net $|$ 0.983, 45.15}]{\includegraphics[width=1.5in]{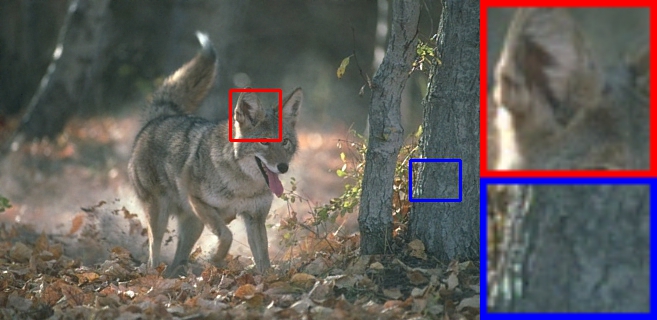}}
	\subfigure[\scriptsize {N2B$_{one}$ $|$ 0.830, 29.56}]{\includegraphics[width=1.5in]{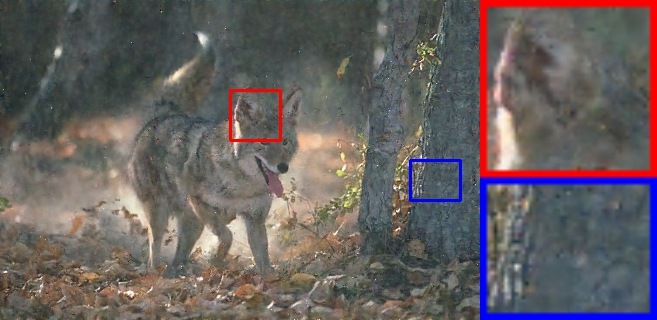}}
	\subfigure[\scriptsize {N2B $|$ 0.967, 38.18}]{\includegraphics[width=1.5in]{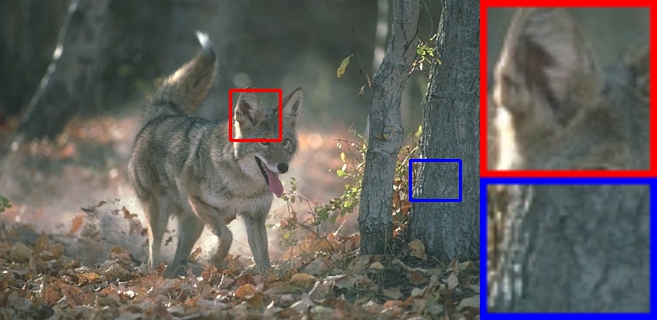}}

	\caption{One visual comparisons for Salt \& pepper denoising, $p=0.15$.}
	\label{fig_salt}
\end{figure*}

\textit{\textbf{Salt \& pepper noise.}}  Salt \& pepper noise can be caused by sharp and sudden disturbances in the image signal. In the disturbed image, each pixel is changed to the maximum or minimum (salt or pepper) with probability $p$. In our training, we vary $p\in [0, 0.3]$. For N2B model, the blurred labels are generated by a median filter with kernel size 31. Quantitative results are shown in Table \ref{table1}. Our N2B results are close to those of CNNs trained with paired data. Subjective results are shown in Figure \ref{fig_salt}. As can be seen, our N2B successfully removes all the noise, and the resulting image is clean and sharp. These experiments prove that our N2B can handle many types of noise. More importantly, our method does not require pre-collected pairs of data. This gives N2B great potential for denoising tasks where paired data is not available.

\begin{figure}[t]
	\centering
	\subfigure[Noise]{\includegraphics[width=1.5in]{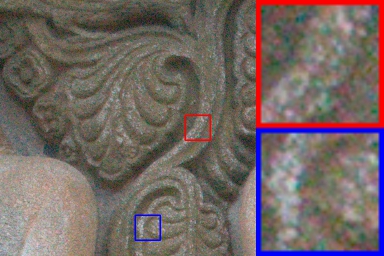}}
	\subfigure[BM3D]{\includegraphics[width=1.5in]{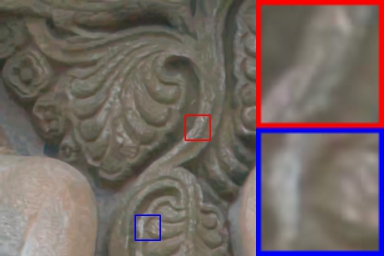}}
	\subfigure[N2V]{\includegraphics[width=1.5in]{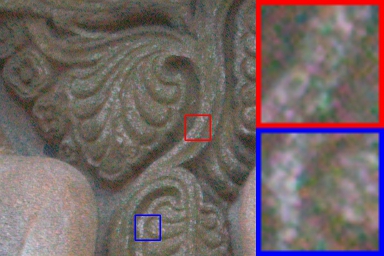}}
	\subfigure[N2B]{\includegraphics[width=1.5in]{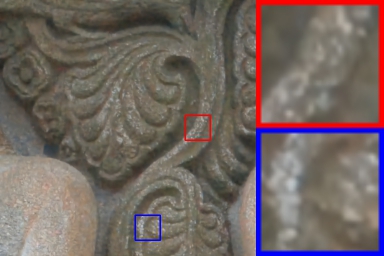}}

	\caption{Denoising results on DND dataset.}
	\label{fig_real}
\end{figure}

\subsection{Real-world noise}
We further evaluate the performance of N2B on a real-world denoising task. We perform our experiments on a benchmark dataset DND \cite{plotz2017benchmarking}. DND contains 50 high-resolution images with realistic noise from 50 scenes. However, it does not provide ground truth and additional pairs of data for training. Therefore, we only compare our method with the self-supervised method N2V and the model-based method BM3D. We crop the 50 noisy images into 1000 image patches with a size $512\times512$. We randomly selected 500 image patches for training and the remaining 500 for testing. Subjective comparisons are presented in Figure \ref{fig_real}. As can be seen, N2V cannot handle the complicated real noise, and BM3D generates over-smoothed images. In contrast, our N2B removes noise cleanly and restores high-quality images.

\subsection{Text removal}
In addition to denoising, N2B can be applied to many image restoration tasks. Similar to denoising, for other image restoration tasks, N2B does not need a priori of the image degradation process, nor does it need pre-collected paired data. Here, we extend N2B to text removal. 

\begin{figure}[t]
	\centering
	\subfigure[\scriptsize{Clean $|$ SSIM, PSNR}]{\includegraphics[width=1.5in]{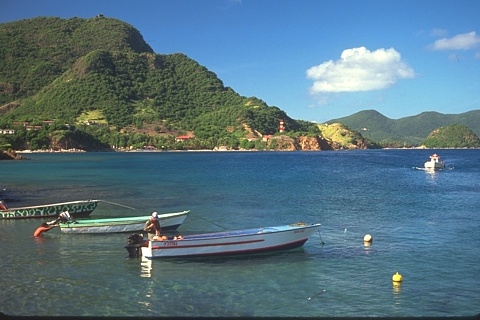}}
	\subfigure[\scriptsize{Input  $|$  0.793, 21.34}]{\includegraphics[width=1.5in]{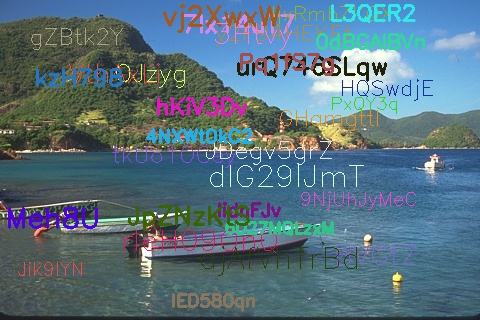}}
	\subfigure[\scriptsize{U-Net $|$ 0.960, 33.71}]{\includegraphics[width=1.5in]{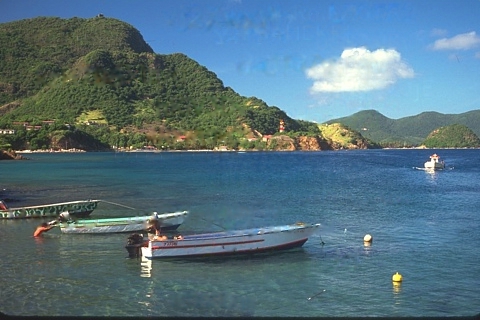}}
	\subfigure[\scriptsize{N2B $|$  0.935, 30.79}]{\includegraphics[width=1.5in]{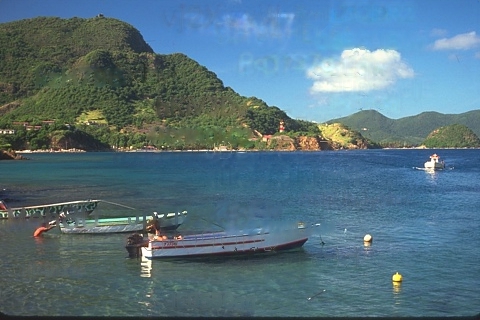}}

	\caption{Text removal example. ($p=0.1$) }
	\label{fig_text}
\end{figure}

We again use the 4744 clean images from \cite{ma2017waterloo} to synthesize text-degraded images. This degradation contains a variety of random strings, which can be random font sizes, random colors, and random locations. During training, each pixel of training samples is degraded with a probability $p\in [0, 0.2]$. When testing, $p$ is set to $0.1$. For text removal, the blurred labels are generated by a median filter with the kernel size 35. Our N2B results are compared with U-Net trained with paired data. We show an example in Figure \ref{fig_text}. In addition, our N2B model yields results 0.964/33.68 dB in terms of SSIM and PSNR for the BSD300 test set, close to the U-Net of 0.978/37.19 dB. These experiments show that N2B is a general method for image restoration.

\subsection{Ablation study}
\begin{figure}[t]
	\centering
	\subfigure[\scriptsize{Clean {\color{white} aaaa} SSIM, PSNR}]{\includegraphics[width=1.9in]{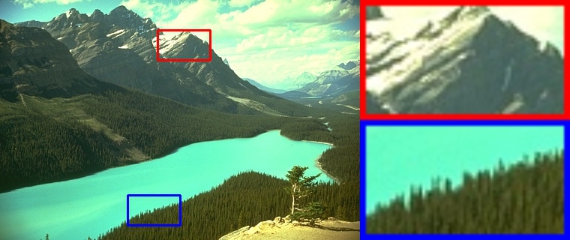}}
	\subfigure[\scriptsize{Input  {\color{white} aaaa} 0.480, 24.27}]{\includegraphics[width=1.9in]{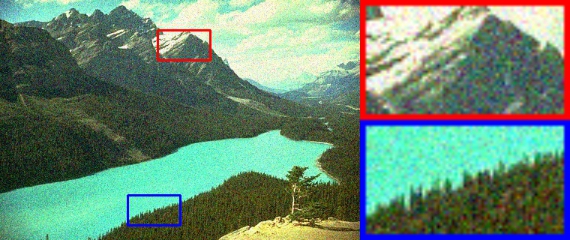}}
	\subfigure[\scriptsize{1 clean image  0.627, 20.12}]{\includegraphics[width=1.9in]{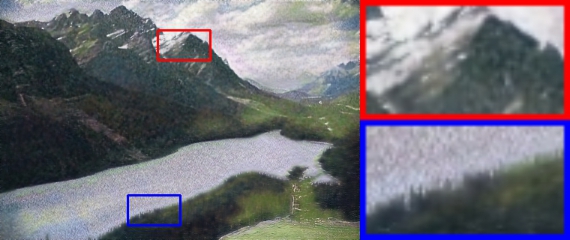}}\\
	\subfigure[\scriptsize{10 clean images 0.768, 23.33}]{\includegraphics[width=1.9in]{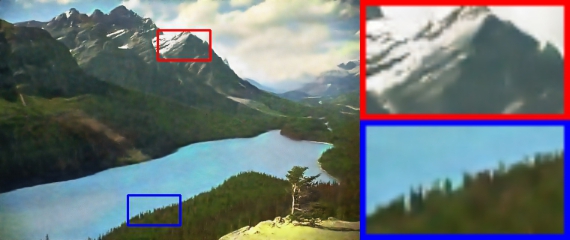}}
	\subfigure[\scriptsize{100 clean images {\color{white} aaaaaaaa} 0.796, 29.55}]{\includegraphics[width=1.9in]{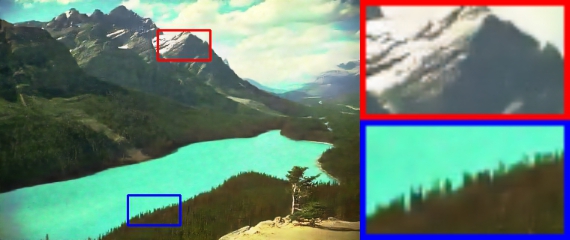}}
	\subfigure[\scriptsize{4000 clean images {\color{white} aaaaaaa} \textbf{0.831}, \textbf{31.43}}]{\includegraphics[width=1.9in]{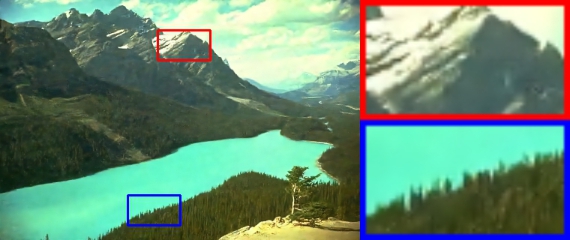}}
	
	\caption{Visual results of N2B models trained using different numbers of clean images.}
	\label{fig_cleannum}
\end{figure}

\textit{\textbf{How many clean images does N2B need?}} As discussed above, N2B can be trained with any number of noisy images. We further explore how many clean images are needed for N2B training. We perform the experiments on Gaussian noise with $\sigma \in [0, 50]$. The N2B model is trained multiple times, using 1, 10, 100, and 4000 clean images, respectively. We report the comparisons in Figure \ref{fig_cleannum} and Table \ref{table_cleannum}. As can be seen, the number of clean images has a significant impact on the performance of N2B. The more clean images available, the better the N2B generalization. Fortunately, in practical applications, a large number of clean images is usually easy to collect.

\begin{table}[t]
	\caption{Quantitative evaluation of N2B models trained with different numbers of clean images.}
	\centering
	\begin{tabular}{|c|c|c|c|c|c|}
		\hline
		\multicolumn{2}{|c|}{Clean images} & 1&10&100&4000\\
		\hline
		
		Gaussian  &SSIM&0.670&0.829&0.858&\textbf{0.873}\\
		\cline{2-6}
		($\sigma=25$)&PSNR&22.36&28.11&29.82&\textbf{30.87}\\
		\hline

	\end{tabular}
	\label{table_cleannum}
\end{table}

\begin{table}[t]
	\caption{Quantitative evaluation of N2B models trained with different image filters.}
	\centering
	\begin{tabular}{|c|p{30pt}|p{30pt}|p{40pt}|p{40pt}|}
		\hline
		\multicolumn{2}{|c|}{Different methods} & mean filter&Bilateral filter&BM3D\\
		\hline
		
		Gaussian &SSIM&0.873&0.873&0.871\\
		\cline{2-5}
		($\sigma=25$)&PSNR&30.53&30.87&30.99\\
		\hline

	\end{tabular}
	\label{table_filter}
\end{table}

\textit{\textbf{Which image filter is better?}} Almost all image filters have a strong denoising ability, but the images they generate are different. Here, we discuss the effects of different filters on N2B. We conduct the experiments on Gaussian noise with $\sigma \in [0, 50]$. 3 different methods are adopted to generate blurred labels, \emph{i.e.} bilateral filter, mean filter with kernel size 31 and BM3D. The image generated by the Bilateral filter still retains obvious edges, while the mean filter removes most of the image details. In addition, BM3D can produce relatively clear and noise-free images (like shown in Figure \ref{fig_lenna}). We report the comparisons in Table \ref{table_filter}. As can be seen, all the methods achieve a similar SSIM, and BM3D has a relatively higher PSNR. However, the performance gap between the different methods is not significant. Therefore, we do not recommend excessive restrictions on the choice of image filters. The only requirement is that the resulting blurred labels must be noise-free.

\begin{figure}[t]
	\centering
	\subfigure[\scriptsize{Clean $|$ SSIM, PSNR}]{\includegraphics[width=1.5in]{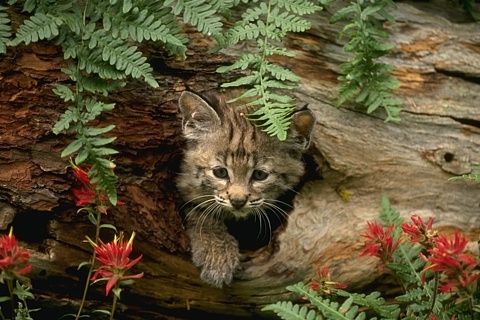}}
	\subfigure[\scriptsize{Input  $|$  0.710, 23.99}]{\includegraphics[width=1.5in]{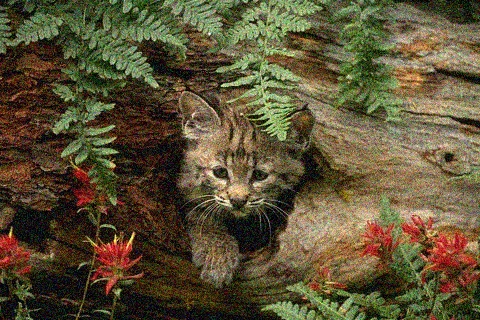}}
	\subfigure[\scriptsize{N2B$_v$ $|$ 0.447, 20.73}]{\includegraphics[width=1.5in]{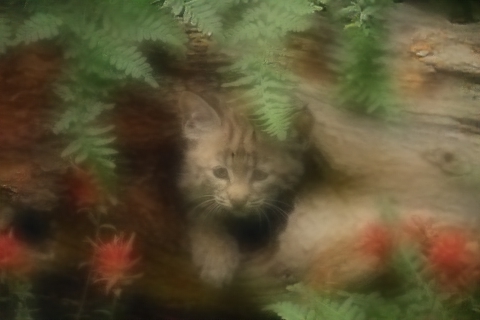}}
	\subfigure[\scriptsize{N2B $|$  0.882, 29.20}]{\includegraphics[width=1.5in]{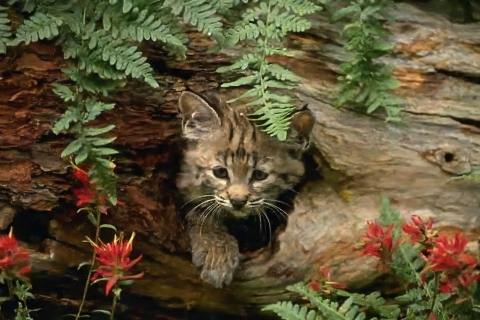}}

	\caption{Gaussian denoising results ($\sigma=25$) of N2B model with/without gradient interruption.}
	\label{fig_ablation}
\end{figure}

\textit{\textbf{Does N2B require gradient interruption?}} To validate the necessity of gradient interruption, we designed a variant of the N2B model called N2B$_v$, which cancels the gradient interruption operation in the convergence stage. Therefore, DnNet is affected by both ``n2b'' Eq. (\ref{eq.n2b3}) and ``n2c'' Eq. (\ref{eq.n2c}) objectives. We conduct the experiments on Gaussian noise with $\sigma \in [0, 50]$. We report the visual comparisons in Figure \ref{fig_ablation}. As can be seen, without gradient interruption, the N2B$_v$ model can also learn to denoise, but loses a lot of image detail. In contrast, gradient interruption can effectively help DnNet focus on the ``n2c'' objective while forcing NENet to finely extract noise. In addition, our N2B gives an average of 0.873/30.87 dB in terms of SSIM and PSNR with a noise level $\sigma=25$, significantly outperform the N2B$_v$ of 0.632/23.10 dB.

\section{Conclusions}

We have proposed Noise2Blur (N2B), a training scheme for blind denoising. N2B enables the training of CNNs without pre-collected paired data. Using only a traditional image filter and some unpaired data, a robust denoising network can be obtained when the N2B model learns to blur the image. We have demonstrated the applicability of N2B in various denoising tasks. In the absence of paired data, N2B still shows encouraging performance. In addition, we have shown the effectiveness of our N2B to other image restoration tasks. We look forward to this method finding more applications in other areas of vision and signal recovery.

{\small
	\bibliographystyle{ieee}
	\bibliography{egbib}
}

\end{document}